\documentclass[final,5p,times,twocolumn]{elsarticle}

\usepackage{amssymb}
\usepackage{amsthm}

\usepackage{graphicx}
\usepackage{dcolumn}
\usepackage{bm}
\usepackage[perpage]{footmisc}
\usepackage{float}
\usepackage{verbatim}
\usepackage{epstopdf}

\journal{Astroparticle Physics}

\begin{document}

\begin{frontmatter}



\title{Measurement and simulation of the muon-induced neutron yield in lead}


\author[address1]{L.~Reichhart\corref{cor1}}
\ead{l.reichhart@sms.ed.ac.uk}
\author[address2]{A.~Lindote}
\author[address3]{D.Yu.~Akimov}
\author[address4]{H.M.~Ara\'{u}jo}
\author[address1]{E.J.~Barnes\fnref{fn1}}
\author[address3]{V.A.~Belov}
\author[address4]{A.~Bewick}
\author[address3]{A.A.~Burenkov}
\author[address2]{V.~Chepel}
\author[address4]{A.~Currie}
\author[address2]{L.~DeViveiros}
\author[address5]{B.~Edwards\fnref{fn2}}
\author[address5]{V.~Francis}
\author[address1,address6]{C.~Ghag}
\author[address1]{A.~Hollingsworth}
\author[address4]{M.~Horn\fnref{fn2}}
\author[address5]{G.E.~Kalmus}
\author[address3]{A.S.~Kobyakin}
\author[address3]{A.G.~Kovalenko}
\author[address7]{V.A.~Kudryavtsev}
\renewcommand{\thefootnote}{\fnsymbol{footnote}}
\author[address4]{V.N.~Lebedenko\footnotemark[2]}
\author[address2]{M.I.~Lopes}
\author[address5]{R.~L\"{u}scher}
\author[address5]{P.~Majewski}
\author[address1]{A.St\,J.~Murphy}
\author[address2]{F.~Neves} 
\author[address5]{S.M.~Paling}
\author[address2]{J.~Pinto da Cunha}
\author[address5]{R.~Preece}
\author[address4]{J.J.~Quenby}
\author[address1]{P.R.~Scovell\fnref{fn3}}
\author[address2]{C.~Silva}
\author[address2]{V.N.~Solovov}
\author[address5]{N.J.T.~Smith\fnref{fn4}}
\author[address5]{P.F.~Smith}
\author[address3]{V.N.~Stekhanov}
\author[address4]{T.J.~Sumner}
\author[address4]{C.~Thorne}
\author[address4]{R.J.~Walker\fnref{fn5}} 

\cortext[cor1]{Corresponding author. Address: School of Physics \& Astronomy, SUPA University of Edinburgh, EH9 3JZ, UK. Tel: +44 131 650 5284.}
\fntext[fn1]{Present address: Department of Physics, Boston University, USA}
\fntext[fn2]{Present address: Department of Physics, Yale University, USA}
\fntext[fn3]{Present address: Department of Physics, University of Oxford, UK}
\fntext[fn4]{Present address: SNOLAB, Lively, Canada}
\fntext[fn5]{Present address: Institut f{\"u}r Experimentelle Kernphysik, Karlsruhe Institute of Technology, Germany}

\address[address1]{School of Physics \& Astronomy, SUPA University of Edinburgh, UK}
\address[address2]{LIP--Coimbra \& Department of Physics of the University of Coimbra, Portugal}
\address[address3]{Institute for Theoretical and Experimental Physics, Moscow, Russia}
\address[address4]{High Energy Physics Group, Blackett Laboratory, Imperial College London, UK}
\address[address5]{Particle Physics Department, STFC Rutherford Appleton Laboratory, Chilton, UK}
\address[address6]{High Energy Physics Group, Department of Physics \& Astronomy, University College London, UK}
\address[address7]{Department of Physics \& Astronomy, University of Sheffield, UK}

\begin{abstract}
A measurement is presented of the neutron production rate in lead by high energy cosmic-ray muons at a depth of 2850~m water equivalent (w.e.) and a mean muon energy of 260~GeV. The measurement exploits the delayed coincidences between muons and the radiative capture of induced neutrons in a highly segmented tonne scale plastic scintillator detector. Detailed Monte Carlo simulations reproduce well the measured capture times and multiplicities and, within the dynamic range of the instrumentation, the spectrum of energy deposits. By comparing measurements with simulations of neutron capture rates a neutron yield in lead of (5.78$^{+0.21}_{-0.28}$)~$\times$~10$^{-3}$~neutrons/muon/(g/cm$^{2}$) has been obtained. Absolute agreement between simulation and data is of order 25$\%$. Consequences for deep underground rare event searches are discussed.

\end{abstract}


\end{frontmatter}

\renewcommand{\thefootnote}{\fnsymbol{footnote}}
\footnotetext[2]{Deceased}


\section{Introduction}

\noindent Rare signal searches, such as those performed for direct dark matter detection (\cite{futuredarkmatter,DDReview,chepelReview} and references therein)  and neutrino-less double beta decay experiments (\cite{doublebeta} and references therein), are typically carried out in deep underground laboratories. The rock over-burden of such facilities removes or dramatically reduces many of the background signals that would be present if the experiments were conducted in surface laboratories. As improved sensitivity is achieved, the need to characterise and mitigate remaining backgrounds becomes ever more important. One of the most problematic backgrounds that still remains is that of cosmic-ray muon-induced neutrons, which may become a limiting factor in some next-generation rare event searches. This specific type of background already shows its impact in current dark matter experiments, with XENON100 reporting it to be the dominant contributor to their nuclear recoil background expectation~\cite{xenon100}.

Neutrons arising from radioactive decays, for example in a fission process or produced in ($\alpha$,n) reactions following \mbox {$\alpha$-decay} of trace contaminations of heavy radio-isotopes, have energies limited to a few MeV. In contrast, neutrons produced through interaction of high energy cosmic-ray muons with matter can reach energies of several GeV. Consequently, while radioactivity neutrons may be effectively controlled by appropriate shielding constructions and selection of radio-pure building materials, removing cosmic-ray induced neutrons is more difficult, with the most effective solution being to go deep underground where the muon flux is reduced by several orders of magnitude compared to that at the surface. Further mitigation of this background involves large muon vetoes, such as instrumented water tanks, to efficiently detect muon tracks far away from the detector.

The neutron production cross-section for high energy muons is very large in high-A materials. Yet, several rare event search projects utilise large amounts of lead to provide shielding against ambient $\gamma$-rays. Thus, the accurate knowledge of the production rate of neutrons by cosmic-ray muons in this material is very important for assessing and planning the capability of these projects, present and future.

A penetrating cosmic-ray muon may produce neutrons via four main processes: 
(i) muon spallation --- muon-nuclear interaction via the exchange of a virtual photon, resulting in nuclear disintegration, 
(ii) muon capture (only dominant for shallow depths, $\lesssim$100~m~w.e.), 
(iii) photo-nuclear interactions in muon-triggered electromagnetic showers, and 
(iv) hadron-production in hadronic cascades initiated by the muon. 
These secondary cascades make up most of the muon-induced neutron production in deep sites. Specifically, neutrons are predominantly created by photo-nuclear interactions of $\gamma$-rays produced in electromagnetic showers, neutron inelastic scattering, pion spallation and pion absorption at rest. The rate of neutron production by direct muon nuclear interaction is significantly smaller than for the other processes listed~\cite{kozlov,Lindote,MUSUN,muonsim,wang}.

The non-trivial task of measuring the cosmic-ray muon induced neutron yield has been pursued by a number of underground experiments (see~\cite{MUSUN,muonsim,wang,mei} for a compilation of such results). Most recently, the KamLAND collaboration has presented muon-induced neutron rates for a number of 
target isotopes~\cite{kamland}. Additional work from other groups is ongoing~\cite{kozlov}. While for low-A targets agreement between the different measurements and simulation toolkits (GEANT4~\cite{geant4}, FLUKA~\cite{fluka1,fluka2}) is reasonable, studies of heavy targets are somewhat controversial and inconsistent~\cite{Kudryavtsev}. Older measurements for Pb targets~\cite{gorshkov,bergamasco}, including beam measurements~\cite{na55}, without Monte Carlo simulations of neutron production, transport and detection, show much larger neutron yields than expected from simulations~\cite{MUSUN,muonsim,wang,marino}. On the contrary, measurements with the veto of the ZEPLIN--II experiment at the Boulby Underground Laboratory showed an over-production in the simulation by $\sim$80$\%$~\cite{muonmeasurement}. 

Here we present a new measurement of the muon-induced neutron yield in lead using the data accrued by a highly segmented anti-coincidence detector installed around the ZEPLIN--III dark matter instrument. The measurement was conducted in parallel to the 319-day long second science run of the experiment in 2010/11.

\section{Experimental apparatus}
\label{sec:experimental}

\noindent The ZEPLIN--III instrument \cite{zeplindesign,zeplinsim} is a dual phase liquid/gas xenon detector, built to observe low energy nuclear recoils resulting from  elastic scattering of weakly interacting massive particles (WIMPs). The final scientific exploitation of the long-running ZEPLIN project at the Boulby Underground Laboratory (at 2850~m~w.e.) achieved cross-section limits for scalar and spin-dependent WIMP--neutron  channels of 3.9~$\times$~10$^{-8}$~pb and 8.0~$\times$~10$^{-3}$~pb near 50~GeV/c$^{2}$ (90$\%$ confidence), respectively~\cite{FSR, SSR}. For the second science run the detector was upgraded with a new array of photomultiplier tubes (PMTs), decreasing internal background sources significantly~\cite{backgrounds}, and a 1-tonne plastic scintillator anti-coincidence detector system (the `veto'), mounted around the main instrument~\cite{veto-design}. Designed for rejecting background events in the WIMP target, the veto also helped to decrease systematic uncertainties in the estimation of background rates due to independent measurements of $\gamma$-ray and neutron rates in the vicinity of the dark matter detector~\cite{veto-performance}. In addition, the veto provided an independent measurement of the high energy cosmic-ray muon flux and its spallation products. The performance of the instrument is well understood through data and validated Monte Carlo simulations~\cite{backgrounds, veto-performance}.

\begin{figure}
\begin{center}
\includegraphics[width=3.0in]{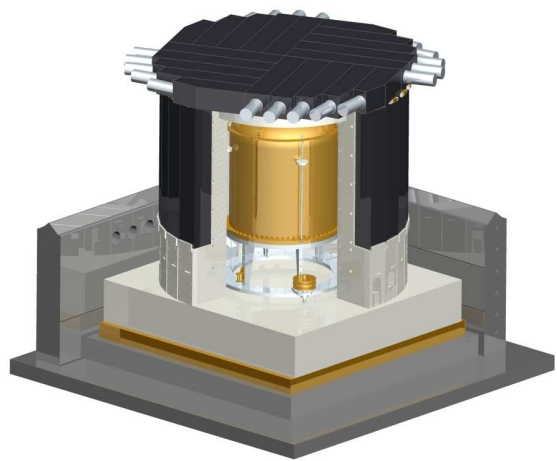}
\caption{(Colour online) CAD rendering of the veto system surrounding the ZEPLIN--III dark matter detection instrument.  The veto barrel consists of 32 vertical Gd-loaded polypropylene pieces (white) surrounded by the same number of active scintillator modules (black), with PMTs housed in cups and recessed into the lower polypropylene structure. The roof of the veto detector is composed of 20 scintillator modules, which are placed on top of a roof plug. The lower polypropylene structure contains no Gd and rests on a copper and lead base. Finally, a lead castle (only the first few lead blocks are shown on the sides facing the back) envelops the entire assembly ($\sim$2.3~m in length and $\sim$2.4~m in height). For display purposes only, a quarter of the scintillator bars from the barrel are not drawn to reveal the ZEPLIN--III detector.}
\label{crosszep}
\end{center}
\end{figure}

The veto system consists of 52 individual plastic scintillator modules (51 were active during the second science run) surrounding a 15~cm thick Gd-loaded polypropylene shielding, which encircles the ZEPLIN--III instrument. This entire structure is then enclosed in a 20~cm thick lead castle. A CAD rendering of the full setup is shown in Fig.~\ref{crosszep}. For a detailed description of the design and performance of each individual component see Ref.~\cite{veto-design}. Here, only a short summary is presented. 

The structure formed by assembling the individual modules can be described by two main geometrical shapes: a circular barrel composed of 32 vertical scintillator bars and a roof constructed from 20 individual scintillator blocks. Each barrel bar has a trapezoidal cross-section with parallel sides of length 15~cm and 12~cm and a height of 15~cm. The length of the barrel scintillators is 1~m. The roof sections are of four different lengths (80, 75, 67 and 51~cm) oriented to form a pseudo-circular shape divided into quadrants and are of rectangular cross-section with side lengths of 15~cm $\times$ 16~cm. The individual detector bars are made from polystyrene-based plastic scintillator UPS-923A (p-terphenyl 2$\%$, POPOP 0.02$\%$), produced by Amcrys-H, Kharkov, Ukraine~\cite{Amcrys-Hwebsite}. A single PMT (ETEL-9302~KB) is optically coupled to one end of each individual scintillator bar.  Additionally, all bars have been wrapped in PTFE sheet of high diffuse reflectivity, and a highly-specular reflective aluminised Mylar film is located at the end opposite to the PMT to increase light collection. 

The polypropylene shielding inside the scintillator construction is loaded with $\sim$0.4$\%$ Gd by weight \cite{veto-performance}. Thus, many neutrons, moderated to thermal energies, undergo radiative capture on $^{157}$Gd due to its very high capture cross-section of 2.4~$\times$~10$^{5}$~barn~\cite{157Gdcrosssec}. This is of great advantage for detecting and identifying  radioactivity neutrons from internal detector components. These are detected with high efficiency through the emission of 3--4~$\gamma$-rays (totalling $\sim$8~MeV) with a mean delay of only $\sim$11~$\mu$s~\cite{veto-performance}. For the more energetic muon-induced neutrons, which are mostly produced externally, a slower capture on hydrogen in the plastic scintillator is expected.

Data were accrued with a dedicated data acquisition system (CAEN model V1724), digitising waveforms with 14-bit resolution, an input range of 2.25 V,  40~MHz bandwidth and a sampling rate of 10~MS/s. The waveforms of recorded events were 320~$\mu$s in length; they were parameterised using a bespoke data reduction software (`RaVen') adapted from that developed for the ZEPLIN--III instrument~\cite{ze3ra}.

The veto detector was operated in `slave' and `master' mode simultaneously. In slave mode the veto acquisition system was triggered by an external signal generated by ZEPLIN--III. The trigger point and timeline lengths were tailored to enable quasi dead time free recording of coincident events. The master mode allowed for independent triggering of the veto system when certain requirements were met. One of these conditions was the sum of simultaneously occurring pulses in the roof modules exceeding a set threshold (summed in a dedicated hardware unit). At this depth, most cosmic-ray muons have an arrival direction which is close to vertical, and thus such a trigger condition provides a high efficiency for detection of cosmic-ray muons, but adds little to the total data storage or rate implied for the experiment.

Critical for a long running experiment is the stability of the detector system over time. Thus, a number of parameters, including electronic gains (measured with the single photoelectron response of the PMTs), coincidence rates, background rates, tagging efficiencies of electron recoil events and environmental parameters, were monitored throughout the course of the experiment. Additionally, a dedicated calibration run was performed on a weekly basis, with a pulsed blue LED, coupled via fibre optic cable to each individual scintillator bar at the end opposite to the PMT. Monitoring of the mean of the single photoelectron peak and of the centroid of the LED-generated peak over the duration of the experiment confirmed the system's stability~\cite{veto-performance}.

\section{Monte Carlo simulations}
\label{sim_general}

\noindent Simulated primary muon energy spectra and angular distributions were obtained by propagation of atmospheric muons from the Earth's surface through an appropriate depth of rock using the MUSIC code~\cite{MUSIC,musicmusun}; this distribution was then sampled with the MUSUN code~\cite{MUSUN,musicmusun}. The energy, momentum, position and charge of each muon was recorded at the point where it intersected the surface of a cuboid fully enclosing the main cavern of the laboratory. The cuboid included an extra 5~m of rock on each side, except for the top which enclosed a total of 7~m of additional rock. The mean energy of the muon distribution was $\sim$260~GeV and 20 million of these muons were generated. The equivalent live-time of the final simulation for the present study amounts to $\sim$3.1~years.

The comprehensive simulation that was developed for the ZEPLIN--III experiment has already been well established in previous studies~\cite{zeplinsim, backgrounds, Leff}. Complementary investigations of the veto detector have also been performed~\cite{veto-design, veto-performance}. This simulation was updated to run with version 9.5 (patch 01) of GEANT4 for this work.

To model the physical processes for this setup the modular physics list {\tt Shielding}, currently recommended for shielding applications at high energies, was implemented.  It uses the Fritiof string model (FTF) and the Bertini cascade (BERT) for the high and low energy ranges (up to 5~GeV), respectively, similar to the {\tt FTFP$\_$BERT} reference list but with different neutron cross-section data (JENDL-HE-2007~\cite{JENDL} up to 3~GeV and evaluated cross-sections~\cite{Barashenkov} above 3~GeV)~\cite{shielding}. Neutron interactions below 20~MeV are described by high-precision data-driven models with data obtained from the ENDF/B-VII library~\cite{endf}. Additionally, thermal scattering off chemically bound atoms was implemented for neutron energies below 4~eV, which is especially important to model thermalisation in the plastics~\cite{thermal}.

Secondary particle production thresholds (`cuts') were set to 0.1~mm for $\gamma$-rays and e$^{-}$/e$^{+}$ which, in lead, translate to $\sim$30~keV and $\sim$250~keV, respectively. This is safely below photo- and electro-nuclear reaction thresholds.

The output generated by the simulation has been designed to recreate that of the experiment, {\em i.e.}~a waveform-like readout with a resolution of 0.1~$\mu$s for all 52 individual channels separately. Thus, direct comparison to data as well as the use of similar analysis cuts for experimental and simulated data is possible.

\section{Event selection}

\noindent During the second science run, it was required that the veto be maximally sensitive to the low energy deposits expected from multiply scattering radioactivity neutrons and $\gamma$-rays. Consequently, bias voltages for each PMT were adjusted to deliver a dynamic range in the region of~1--70~photoelectrons (phe), corresponding to approximately 20--1300 keV at the far end of the scintillator. A minimum-ionising muon crossing the full thickness of a scintillator bar deposits at least $\sim$20~MeV and, thus, muon signals, and a greater number of MeV energy deposits from ambient \mbox{$\gamma$-ray} background, result in heavily saturated pulses. Given a single range data acquisition, recording of non-saturated muon events simultaneously with the signal expected from captured neutrons would not be possible. Selection of muons from this data set is therefore non-trivial, but can be achieved by searching for coincident saturated signals in roof and barrel scintillators, due to the optical separation of the modules. 

\begin{figure}
\begin{center}
\includegraphics[width=0.45\textwidth]{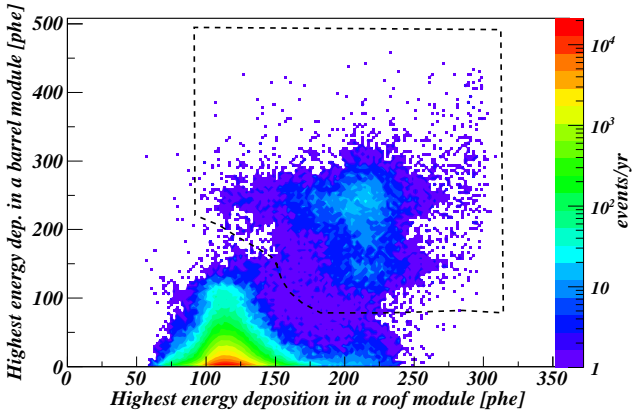}
\caption{(Colour online) Highest energy deposition observed in one of the roof scintillator modules versus the time-coincident highest energy deposition in a barrel scintillator module. The dashed line indicates the graphical cut used to select the muons. Note that energy depositions $\gtrsim$100~phe are saturated.}
\label{data_muon_select}
\end{center}
\end{figure}

Figure~\ref{data_muon_select} shows the greatest energy deposition observed in a roof module plotted against the largest corresponding (coincident) signal in a barrel module for each event, occurring within $\pm$0.2~$\mu$s around the trigger point. This is similar to the prompt coincidence window used for tagging $\gamma$-ray events in \mbox{ZEPLIN--III}~\cite{veto-performance}. A well separated population is observed, with a graphical selection criterion indicated. Here, the photoelectron scale is defined by using a constant conversion factor between the pulse area and the pulse height parameter, which was utilised for the single photoelectron calibration. Given that the pulse area is less affected by saturation than the pulse height (due to the abrupt cutoff in the latter), the impact of saturation can be pushed to higher energies ($\gtrsim$100~phe), and so improve separation of event populations.

\begin{figure}
\begin{center}
\includegraphics[width=0.45\textwidth]{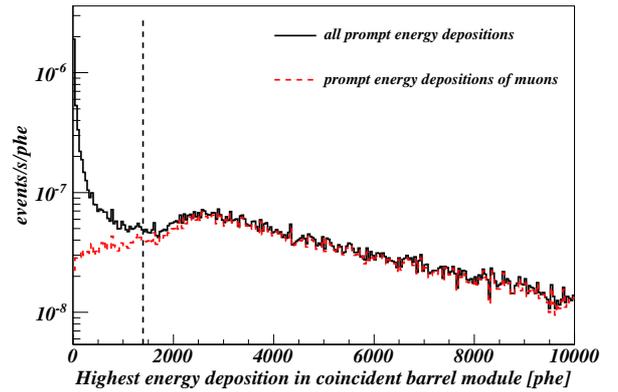}
\caption{(Colour online) The plot shows the highest energy deposition observed in a barrel scintillator module when measured in coincidence with a roof module for all prompt energy depositions in the simulation (black solid). The red dashed spectrum shows the same but for muon hits only, {\em i.e.}~the muon crosses both the roof and the coincident barrel module. The cut used to select muon events is indicated by the thick dashed vertical line. $\sim$88$\%$ of muons with a roof -- barrel coincidence have energies above this threshold.}
\label{muon_eff}
\end{center}
\end{figure}

Selection of muons in the simulation followed a very similar procedure. Firstly, events with a minimum energy deposition observed from the summed signal of the veto roof, analogous to the trigger function of the veto detector, were selected. Additionally, as in the data, a cut on time coincidence (0-0.4~$\mu$s) between roof and barrel was applied. In Fig.~\ref{muon_eff}, the Monte Carlo data are plotted as a function of the largest energy deposition in the (coincident) barrel module only. Separate curves are shown for all events satisfying the coincidence condition, and for only those events corresponding to energy depositions directly resulting from muon traversal of scintillator modules. The difference between the two curves is predominantly due to the energy depositions from particles generated in showers as muons pass nearby. A simple cut at the position indicated by the dashed line selects a population which is composed of~$\sim$93$\%$ muon energy depositions with the required coincidence, {\em i.e}~at least one roof and one barrel module firing within the defined coincidence window.

\begin{figure}[h]
\begin{center}
\includegraphics[width=0.5\textwidth]{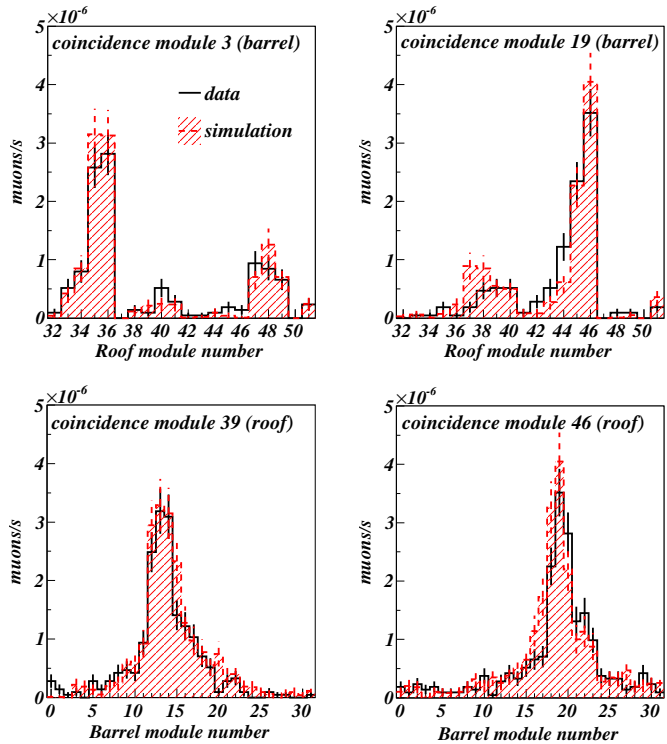}
\caption{(Colour online) Sample of coincident channels from two barrel slabs (top, module 3 and 19) and two roof slabs (bottom, module 39 and 46) with all modules (channels) of the roof and the barrel, respectively. The simulation, scaled to the total muon rate observed in the data, is shown by the red dashed hatched histogram in comparison to the data (black histogram).}
\label{coinc_channels}
\end{center}
\end{figure}

Confirmation that the identified region in the experimental data corresponds to the muon event region in the Monte Carlo is provided by comparing the event distributions between pairs of roof modules and barrel modules (scaled to the overall observed muon rate as measured from the experimental data), as shown in Fig.~\ref{coinc_channels}. Here, the two upper panels show the distribution of roof modules (numbered 32--51) registering a coincidence with a specific barrel module (modules 3 and 19, as indicated). Similarly, the lower panels show which barrel modules (numbered 0--31) are in coincidence with which roof modules (39 and 46). The inactive module is one of the central roof scintillator bars (number 50) featuring a length of 80~cm. The combination of the relative orientations of the modules with respect to each other, their individual response functions, and the asymmetric impact of the surrounding laboratory geometry, results in a complex distribution of coincidences between modules. However, the Monte Carlo reproduces the experimental data reasonably well (the average reduced $\chi^{2}$ value of these 51 coincidence contributions is $\sim$1.9), confirming that the selected experimental data correspond to cosmic-ray muon events.

An overall efficiency for pure muon events, including previously mentioned effects and the (geometric) requirement for coincidence between barrel and roof is 36.8$\pm$0.6$\%$, where the error includes uncertainties due to the precise choice of the location of the selection cuts.

A total number of 7979 muons was selected from the full dataset translating to a rate of 32.3$\pm$0.4~muons/day.
By comparing the measured rate to the Monte Carlo prediction, using the normalised flux through a sphere in the simulation in a similar way to \cite{muonmeasurement}, we deduce a muon flux of (3.75$\pm$0.09)~$\times$~10$^{-8}$~muons/s/cm$^{2}$. This result is in excellent agreement with the last reported value for the muon flux in the Boulby Underground Laboratory of (3.79$\pm$0.15)~$\times$~10$^{-8}$~muons/s/cm$^{2}$ \cite{muonmeasurement}, measured in the cavern hosting both the ZEPLIN--II and ZEPLIN--III detectors, and $\sim$8$\%$ lower than the value obtained for another cavern in Boulby reported in~\cite{Robinson}.

\section{Muon-induced neutron yield}

\noindent The vast majority of detected neutrons produced by muons in this set-up originates in the $\sim$60-tonne lead shield, which protects the experiment from ambient $\gamma$-rays. To determine the muon-induced neutron yield in lead from the present data we count the number of neutrons captured in the veto following a recorded muon event. This is compared with simulations performed using the same analysis cuts. We note that a data set with single photoelectron resolution is a real asset: at the expense of a small increase in background rate, the low threshold analysis increases the number of detected neutrons substantially in comparison with previous works.

\subsection{Experiment}
\label{exp_neutrons}

\begin{figure}
\begin{center}
\includegraphics[width=0.45\textwidth]{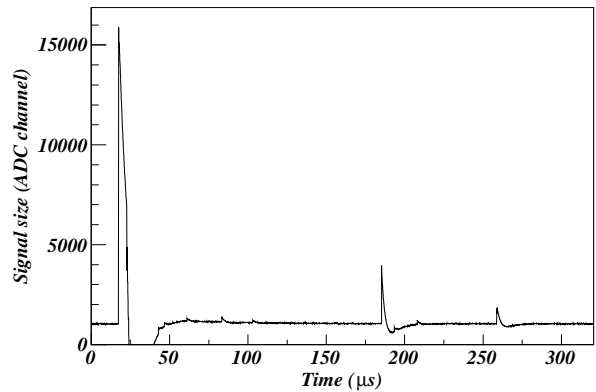}
\caption{Sample waveform showing a highly saturated pulse ($\sim$20~$\mu$s) from an energy deposition of a muon passing through a roof scintillator bar. The saturated pulse is followed by significant overshoot. Single photoelectron-like pulses are visible where the pulse overshoot starts to recover (between 40--50~$\mu$s). These are suppressed by the pulse-finding algorithm as these pulses are below the baseline of the waveform. At $\sim$185~$\mu$s the delayed signal from an accepted muon-induced neutron event (this particular capture signal was observed in 5 scintillator bars simultaneously) is shown and another one at $\sim$260~$\mu$s. In this case the signal, with a size of 4~phe, is part of a channel multiplicity 2 event.}
\label{waveform}
\end{center}
\end{figure}

\begin{figure}
\begin{center}
\includegraphics[width=0.45\textwidth]{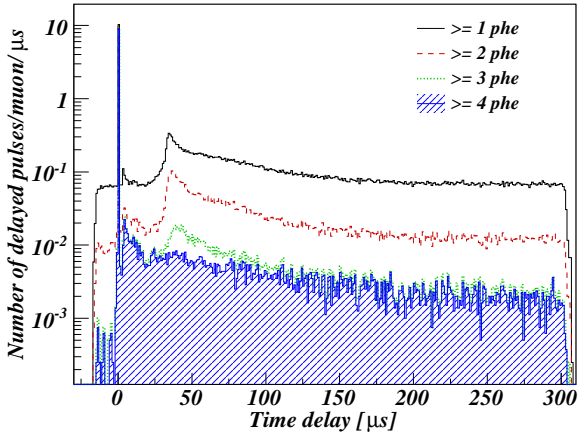}
\caption{(Colour online) Time delay distribution for all recorded pulses above given thresholds relative to the muon.}
\label{timeline}
\end{center}
\end{figure}

\noindent As described previously, neutrons are identified through signals occurring in one or more of the 51 scintillators as a result of the $\gamma$-rays emitted following their capture. These signals are delayed relative to the muon's passage due to the time for thermalisation and capture to occur. Ideally, the data would be searched for the signatures of neutron captures over the entire period in which these signals may arrive. However, the PMT response to a large energy deposition is such that the timelines become at first heavily saturated, and then exhibit a large signal overshoot. For extreme energy depositions the overshoots persisted for up to 40~$\mu$s. A sample waveform of a heavily saturated signal from an energy deposition of a muon passing through a roof scintillator bar is given in Fig.~\ref{waveform}. The effect of these `dead' waveform periods can be seen in Fig.~\ref{timeline}, showing a significantly reduced pulse rate for the first $\sim$40~$\mu$s after the muon trigger. Thus, the timeline for detecting delayed neutrons was restricted to the region of 40--300~$\mu$s relative to the observed muon. An efficiency of $\sim$47$\%$ was retained from this timeline selection cut (calculated from simulations). Furthermore, the maximum number of recorded pulses was restricted to 300 entries per event (an equivalent cut was implemented in the analysis of the simulation). The impact of this restriction is discussed in Section~\ref{neutron_sim}.

For the detection of muon-induced neutrons, as compared to internally-generated radioactivity neutrons, one expects an increased importance of neutrons capturing on hydrogen. In analysing veto data to support the dark matter search, neutrons will have scattered within the \mbox{ZEPLIN--III} instrument and thus have a high geometrical probability of being captured in the Gd-loaded polypropylene shielding immediately surrounding the target. Most muon-induced neutrons come from outside of the setup and will more likely be captured in the hydrocarbon scintillator material surrounding the Gd-loaded shielding. A single $\sim$2.2~MeV $\gamma$-ray is emitted following capture on hydrogen, and therefore signals observed in a single scintillator module are more likely to occur -- in contrast to the several $\gamma$-ray signature from Gd capture, which can be recorded simultaneously in several scintillator modules. Due to the relatively long capture times of neutrons on hydrogen in comparison to captures on gadolinium, the rejection of the first 40~$\mu$s of the waveforms reduces the probability of detection with single scintillator signals by only $\sim$26$\%$.

\begin{figure}
\begin{center}
\includegraphics[width=0.45\textwidth]{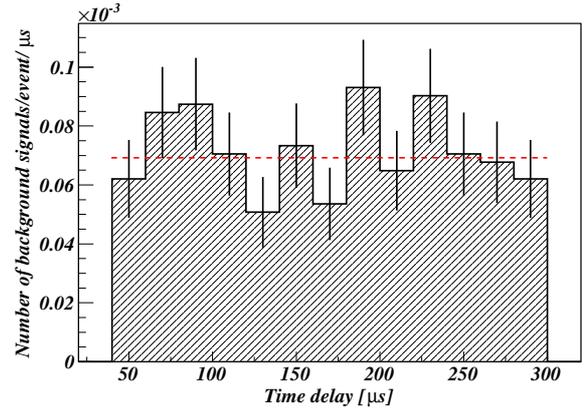}
\caption{(Colour online) Time delay distribution of channel multiplicity one events (threshold of $\geqslant$10~phe) in ZEPLIN--III coincident background data. The dashed (red) line indicates the constant fit to the background.}
\label{const_background}
\end{center}
\end{figure}

Single scintillator events are more exposed to backgrounds, and careful consideration of thresholds and a good knowledge of those backgrounds are required. Since available statistics of the limited pre-trigger timeline fraction are very scarce, an additional data set from the same run with similar trigger conditions was used to estimate the background correctly. A dataset of synchronised (with ZEPLIN--III) `slave' triggered veto events (see Section~\ref{sec:experimental}) was considered to calculate the background. As shown from the analysis of the dark matter search data, these events are fully consistent with $\gamma$-ray background~\cite{SSR}. If tagged by the veto, a prompt signal occurs within a time window of 0.4~$\mu$s~\cite{veto-performance}. Any signals recorded in the waveforms of the veto some $\mu$s away from the trigger time are due to uncorrelated background events only. Optimisation of the number of neutron captures observed in the muon triggered data, with respect to the number of false events due to background, results in a threshold of $\geqslant$10~phe being chosen for single scintillator events (the present results were shown to be largely insensitive to the precise threshold). Figure~\ref{const_background} shows the rate of background events with a threshold of $\geqslant$10~phe applied. The rate is approximately constant, {\em i.e.}~it is independent of the time since the trigger occurred. 

Following the methodology used in the analysis of the dark matter search data~\cite{SSR}, coincident signals in multiple scintillators can also be searched for, detecting multiple scatters and $\gamma$-rays following neutron capture on gadolinium at later times. Coincidences are defined as occurring within $\pm$0.2~$\mu$s of each other. To optimise efficiency, different signal size thresholds have been required depending on the number of signals in coincidence, balanced against the rate of false signals arising from non-neutron related sources (background and induced noise). Noise from the PMTs can be intrinsic, {\em i.e.}~from thermionic emission and internal radioactive decays, or directly induced. Especially after larger signals, such as resulting from muons, positive ions generated from ionisation of residual gases in the PMTs lead to secondary signals, creating afterpulses at short time scales of up to several $\mu$s dependent on the ion transit time (see~\cite{Birks,Campbell} and references therein). In the present data afterpulses of small amplitudes are suppressed at short decay times due to the large pulse overshoots observed following a muon energy deposit (see Fig.~\ref{timeline}). A second noise component observed at longer time scales, visible in Fig.~\ref{timeline} between $\sim$30--50~$\mu$s after the start of the muon signal, with sizes of 4~phe and below, may be attributed to the organic scintillator. Luminescence with long time constants is expected from phosphorescence and delayed fluorescence processes in the plastic scintillator (see {\em e.g.}~\cite{Birks,Kawada,Marvin}).

These additional signals could lead to false coincidences between scintillators, generating spurious neutron detections. Based on the event rates, the probability of false coincidences can be calculated. It was found that for neutron capture events with a channel multiplicity of two, {\em i.e.}~two scintillator bars firing within $\pm$0.2~$\mu$s of each other, a signal size requirement of threshold $\geqslant$4~phe (in each pulse) was sufficient to remove afterpulses. For three-fold coincidences between scintillators, a threshold of $\geqslant$2~phe per signal was found to be appropriate, and for four or more scintillators, a threshold at the level of a single photoelectron was sufficient. For consistency, a global requirement was set that regardless of the number of scintillators fired in coincidence, all events must have a total signal size of at least 8~phe. Despite the lower threshold for multiple scintillator events, accidental rates arising from background are a lot smaller due to the required coincidence of pulses. The same dataset used earlier to estimate the background rate in the single scintillator case has thence been utilised to calculate the contribution from background to the yields of neutron captures found from the multiple scintillator requirements. Background rates found are at the level of statistical uncertainties.

Table~\ref{data_neutron_rates} summarises the results. Each instance in which the designated criteria were met is interpreted as indicating a neutron capture. Most muon-induced neutron captures are observed through events seen in single scintillators only, despite the higher threshold required. However, a significant number also generate energy depositions observed in coincidence in several scintillators. Overall, a mean of 0.346$\pm$0.007~neutrons (including background corrections) are observed for every muon detected. 

\begin{table}
  \centering 
  \caption{Measured number of neutrons per muon from the data in comparison to neutron rates extracted from simulations using the same requirements and cuts as in the experimental data analysis. Background rates, for correction of the data, are listed individually for the different channel multiplicities, with their required thresholds detailed in the text. The errors given for the data are the sum of statistical errors and the rate coming from random accidental coincidences of pulses calculated from the average observed pulse rate for a given threshold. Errors of simulated rates are statistical only.}
    \vspace{2mm}
  \label{data_neutron_rates}
   \begin{tabular}{p{1.0cm} p{1.45cm} p{1.45cm}p{1.45cm}|p{1.45cm}}
  \hline
  \multicolumn{5}{c}{\hspace{33mm}Data\hspace{29mm}Simulation}\\
  \hline
 Channel mult. & Events/ muon & Background rate & n/muon (bkg.corr.) & n/muon\\
 \hline
 \hline
 1 & 0.216(5) & 0.019(1) & 0.197(5) & 0.145(2)\\
 2 & 0.088(3) & 0.0049(5) & 0.083(3) & 0.076(1)\\
 3 & 0.039(2) & 0.0019(3) & 0.037(2) & 0.0321(9)\\
 $\geqslant$4 & 0.029(2) & 0.0008(2) & 0.028(2) & 0.0231(8)\\
 \hline
 \hline
 Total & 0.372(7) & 0.026(1) & 0.346(7) & 0.275(3)\\
 \hline
\end{tabular}
\end{table}

\subsection{Comparison with simulations}
\label{neutron_sim}

\noindent Monte Carlo simulations of the experiment have been performed as described in Section~\ref{sim_general}. The dimensions and parameters of the apparatus have been previously measured and documented~\cite{veto-design,Leff,qfpaper}, including signal gains and attenuation lengths of the scintillators so that photoelectron spectra can be generated. This allows Monte Carlo pseudo-data to be analysed using identical routines as used for the real data. The overall agreement on the rate of detected neutrons between data and simulation obtained in this work is good, at the level of 25$\%$. For the initial discussion the total number of neutrons has been normalised to the data. 

\begin{figure}
\begin{center}
\includegraphics[width=0.45\textwidth]{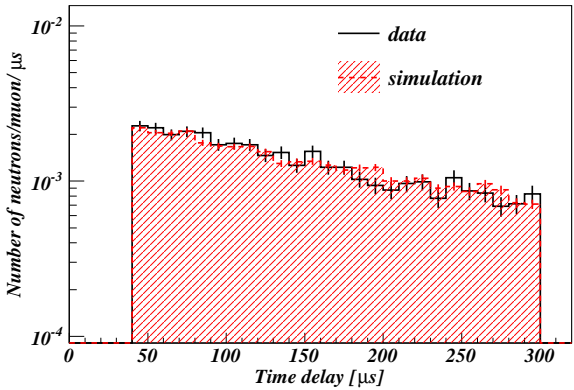}
\caption{(Colour online) Time delay distribution of detected captured neutrons from experimental data (black solid histogram) and simulations (red dashed hatched histogram). The constant background has been subtracted from the data histogram. Results from simulation are normalised to the total number of neutrons observed in the data.}
\label{neutron_time_delay}
\end{center}
\end{figure}

\begin{figure}
\begin{center}
\includegraphics[width=0.45\textwidth]{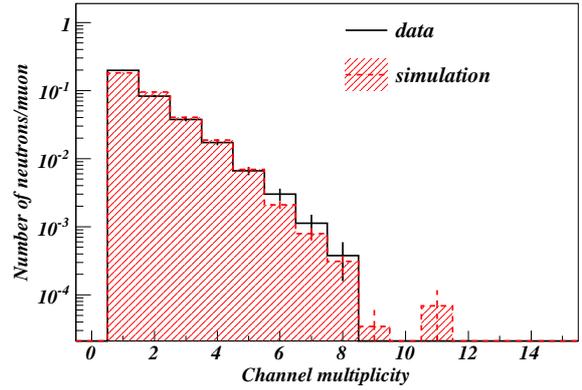}
\caption{(Colour online) Comparison of channel (scintillator module) multiplicities per detected neutron in the data (black solid) to simulations (red dashed hatched histogram). The data are background corrected according to Table~\ref{data_neutron_rates}. Results from simulation are normalised to the total number of neutrons observed in the data.}
\label{neutron_channel_mult}
\end{center}
\end{figure}

Figure~\ref{neutron_time_delay} compares the time delay distributions for detected neutrons from data (solid black) and simulation (red dashed). Excellent agreement between the two distributions is found. Moreover, the module (channel) multiplicity per neutron event can be used as an additional consistency check, beyond the initial muon identification, taking advantage of the segmented nature of the detector. Figure~\ref{neutron_channel_mult} shows the number of channels with coincident signals involved in each individual neutron event. The data are corrected for the contributions from background coincidences, as given in Table~\ref{data_neutron_rates}. Again, excellent agreement, over the full range of channel multiplicities, is demonstrated.

\begin{figure}
\begin{center}
\includegraphics[width=0.45\textwidth]{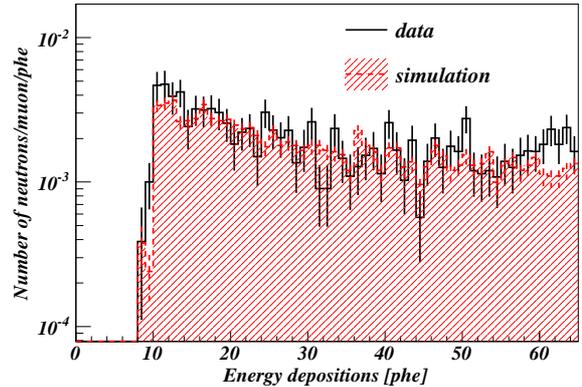}
\caption{(Colour online) Energy depositions of detected neutrons from background corrected data (black solid) and simulations (red dashed hatched histogram) below the saturation point in the data, {\em i.e.}~the energy scale is given in absolute number of photoelectrons (1~phe~$\simeq$~20~keV). Results from simulation are normalised to the total number of neutrons observed in the data.}
\label{neutron_edep}
\end{center}
\end{figure}

In Fig.~\ref{neutron_edep} the energy depositions associated with the observed (captured) neutrons are given in the region before the onset of saturation, with excellent agreement between simulation and data obtained. Although the energy calibration is only known to within 10$\%$ due to, amongst other factors, the saturation of the data (see also previous studies with the same instrument~\cite{veto-performance,qfpaper}), tests in varying the energy scale by this amount resulted in only small neutron rate differences and are considered in the systematic error of the simulated rate.

\begin{figure}
\begin{center}
\includegraphics[width=0.45\textwidth]{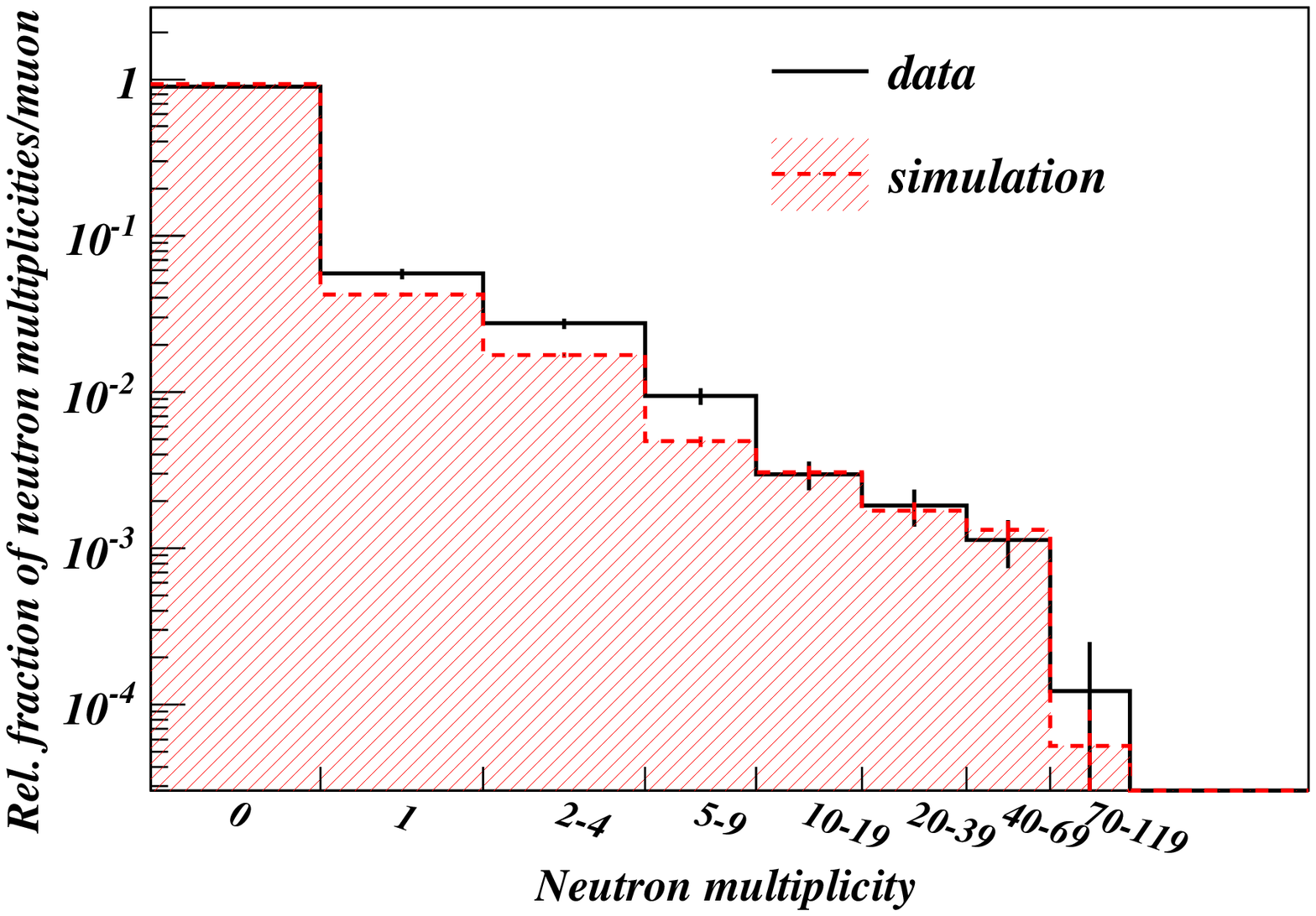}
\caption{(Colour online) Relative fraction of neutron multiplicities per muon, {\em i.e.}~the number of delayed signals observed after a muon trigger in the defined time window, for background corrected data (black solid) and simulation (red dashed hatched histogram) normalised to the total number of observed muons in each case. 66$\%$ of neutron capture signals associated with muon events which registered a single detected neutron only are observed in a single scintillator.}
\label{neutron_neutron_mult}
\end{center}
\end{figure}

A muon may produce more than one fast neutron in a cascade, resulting in several neutron capture signals at different times and in different locations in the veto. Figure~\ref{neutron_neutron_mult} shows the relative fraction of observed neutrons per muon for data and simulation. When exploring neutron multiplicities, rather than scaling the simulation to the total number of neutrons observed in the data, a simple normalisation to the number of detected muons has been applied. Background corrections assume even distribution of background events. As such, most non-neutron signals occur in one of the empty waveforms following a muon trigger, making up almost 90$\%$ of all observed muon events. Generally, good agreement is observed.

When scaled to the number of neutrons detected, the simulations reproduce well the time distributions, the energy depositions and the number of scintillators involved in each event. The absolute numbers of neutrons expected to be observed per muon, as determined from the simulation for each individual channel multiplicity, are also summarised in Table~\ref{data_neutron_rates}, showing an overall reduced neutron rate from simulations of $\sim$20$\%$ ({\em i.e.}~the total yield from the data exceeds the simulation by $\sim$26$\%$). Discrepancies are largest for single scintillator events. At higher multiplicities absolute agreement between simulation and data is of order 10--20$\%$ (cf.~$\sim$36$\%$ for single scintillator events).

The expected total muon-induced neutron rate calculated from simulations is 0.275$\pm$0.003~(stat.)~$^{+0.004}_{-0.007}$~(syst.)~neutrons/muon. Systematic errors are calculated from the variability in the energy calibration.

To assess the greater discrepancy for single channel neutron events between data and simulation, tests in raising the detection threshold and limiting the time window to search for neutron capture signals (further away from the trigger) were performed. No significant differences are observed. Additionally, exclusion of the two channels with the highest energy depositions from the traversing muon (used prior to the muon-induced neutron analysis for the initial selection of muon events) from the final analysis reduces the number of neutron captures by almost the same factor in data and simulations. For neutron captures giving a signal in one scintillator only, the difference in the reduction factors is practically the same as for higher channel multiplicities. For the whole sample of neutron captures the slight difference in the reduction factors between data and simulations introduces a systematic uncertainty of 4$\%$. Thus, the overall detected muon-induced neutron rate from experimental data results in 0.346$\pm$0.007~(stat.)~$^{+0.000}_{-0.014}$~(syst.)~neutrons/muon.

As previously mentioned, a restriction on the maximum number of recorded pulses was applied to the data, and similarly in the analysis of the simulation. Importantly, the number of selected muon events, in both data and simulation, affected by this limitation is in excellent agreement, further supporting the performance of the Monte Carlo simulation. In the data 11 events were found which were affected by this cut; the number of events associated with more than 300 pulses in the simulation (scaled to the data) amounts to 11$\pm$2. When including all energy depositions in the simulation a higher absolute neutron rate is observed ($\sim$30$\%$). This increase is associated with only a few muon events (approximately 1 in 1700) featuring exceptionally high neutron multiplicities. It is worth noting that these high multiplicity events are less significant for dark matter searches due to the generally high veto efficiencies expected for these.

Table~\ref{neutrons_produced} shows the relative production of neutrons in different materials for all neutrons generated in the simulation and for detected neutrons only. As expected, nearly all neutrons are produced in the rock cavern of the underground laboratory, reflecting that the simulation included sufficient volume to remove edge effects. Importantly, less than 1.5$\%$ of detected neutrons are produced in the rock, confirming the effectiveness of the shielding setup of the ZEPLIN--III detector. On the other hand, the lead component of the shielding enclosure provides an effective target for neutron production by high energy comic-ray muons, with $\sim$95$\%$ of all neutrons created there. 

Table~\ref{neutron_capture_element} lists the specific elements involved in the capture of the neutrons, both for detected neutrons and for all the neutrons in the simulation. As previously mentioned, the vast majority of detected muon-induced neutrons are captured on hydrogen, emphasising the importance of measuring the single $\sim$2.2~MeV $\gamma$-ray from this process. The captures on Gd amount to $\sim$7.0$\%$ for this configuration at these neutron energies (being much more effective for internal radioactivity neutrons due to the detector geometry described in Section~\ref{sec:experimental}).

\begin{table}
  \centering 
   \caption{Fractions of neutrons produced in different materials for all generated neutrons in the simulation and for detected neutrons only.}
       \vspace{2mm}
   \label{neutrons_produced}
  \begin{tabular}{p{2.5cm} p{2.5cm}  p{2.5cm}}
  \hline
  \multicolumn{3}{c}{\hspace{23mm}Production material of}\\
 Material & all neutrons & detected neutrons\\
 \hline
 \hline
 Lead & 0.2$\%$ & 95.0$\%$\\
 Rock & 99.8$\%$ & 1.4$\%$\\
 Steel & - & 1.2$\%$\\
 C$_{8}$H$_{8}$ & - & 0.9$\%$\\
 Copper & - & 0.8$\%$\\
 CH$_{2}$ & - & 0.5$\%$\\
 Gd-epoxy & - & 0.1$\%$\\
 Liquid Xe & - & 0.1$\%$\\
 \hline
\end{tabular}
\end{table}

\begin{table}
  \centering 
   \caption{Fractions of neutrons captured on different elements for all and for detected neutrons only.}
       \vspace{2mm}
   \label{neutron_capture_element}
  \begin{tabular}{p{2.5cm} p{2.5cm}  p{2.5cm}}
  \hline
  \multicolumn{3}{c}{\hspace{23mm}Capture element of}\\
 Element & all neutrons & detected neutrons\\
 \hline
 \hline
 H & - & 71.1$\%$\\
 Fe & - & 11.5$\%$\\
 Cl & 94$\%$ & 7.0$\%$\\
 Gd & - & 7.0$\%$\\
 Pb & - & 1.3$\%$\\
 C & - & 1.1$\%$\\
 Cu & - & 0.6$\%$\\
 Na & 6$\%$ & 0.2$\%$\\
 Mn & - & 0.2$\%$\\
 \hline
\end{tabular}
\end{table}

\subsection{Muon-induced neutron yield in lead}

\noindent As shown in Table~\ref{neutrons_produced}, the detected neutrons have predominantly been produced in lead. Thus, the observed neutron rate may be used to derive an absolute neutron production yield in this material. The methodology used follows that of Refs.~\cite{Lindote,muonmeasurement}, and is essentially to scale an idealised simulation of neutron production by a mono-energetic beam of muons in pure lead by the ratio in rate observed between the present data and the full detector simulation (assuming that the fraction of detected neutrons produced in lead ($\sim$95$\%$) is well described by the simulation). 

\begin{figure}
\begin{center}
\includegraphics[width=0.36\textwidth]{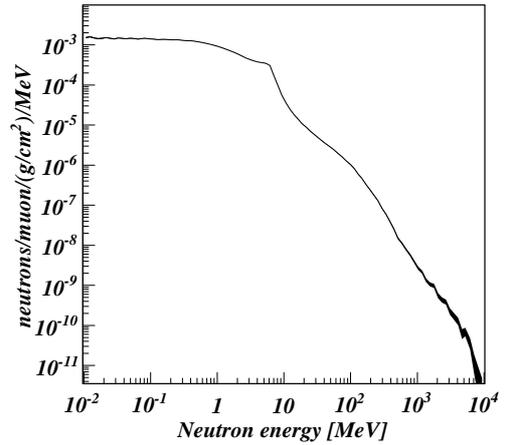}
\caption{Differential energy spectrum of muon-induced neutrons produced in lead from $\mu^{-}$ of 260 ~GeV.}
\label{neutron_yield_in_lead}
\end{center}
\end{figure}

The simulation of a mono-energetic muon beam in lead was conducted as follows. Neutron production was recorded for a mono-energetic 260~GeV $\mu^{-}$~beam (mean muon energy at Boulby), incident on the centre of a lead block of 3200~g/cm$^{2}$ thickness. Figure~\ref{neutron_yield_in_lead} shows the differential energy spectrum of neutrons produced. Only neutrons from the central half length of the lead block were considered to avoid surface/edge effects. To prevent double counting in neutron inelastic processes, the first neutron produced in each reaction was dismissed independently of its energy. As summarised in Table~\ref{neutron_yield_table}, a production rate of (4.594$\pm$0.004)~$\times$~10$^{-3}$~neutrons/muon/(g/cm$^{2}$) was obtained for the physics list and version of GEANT4 used throughout this work ({\tt Shielding} with version 9.5). However, since the experimental muon-induced neutron rate was found to be a factor of 1.26$\pm$0.03~(stat.)~$^{+0.03}_{-0.05}$~(syst.) higher, our results suggest a true production rate by 260~GeV muons of (5.78$\pm$0.13~(stat.)~$^{+0.16}_{-0.25}$~(syst.))~$\times$~10$^{-3}$~neutrons/muon/(g/cm$^{2}$), assuming neutron transport and detection are modelled accurately.

In this analysis, uncorrelated arrival of muons is assumed, as opposed to muon bundles produced together by primary cosmic-rays in the atmosphere. A study based on a simple approximation to find the survival probability of muons at a given depth showed the effect to be negligible and the error to be very small for the measurement performed with the \mbox{ZEPLIN--III} veto detector.

A similar comparison between simulation and experiment was performed for the ZEPLIN-II anti-coincidence system~\cite{Lindote,muonmeasurement}. In that work a muon-induced neutron yield in lead of (1.31$\pm$0.06)~$\times$~10$^{-3}$~neutrons/muon/(g/cm$^{2}$) was reported. Here we have revisited the simulation, now using GEANT4 version 9.5 and the {\tt Shielding} physics list, including thermal scattering cross-sections, and a significantly larger sample of primary muons. This resulted in a new estimate for the neutron yield in lead of (3.4$\pm$0.1)~$\times$~10$^{-3}$~neutrons/muon/(g/cm$^{2}$) in that setup. While it is clear that a significant contribution of the newly obtained ZEPLIN--II yield comes from the updated simulation, there remains a significant discrepancy with the present result. One possible explanation for this is that the angular distribution of emitted neutrons may not be accurately modelled. The ZEPLIN-III veto scintillators are predominantly sensitive to neutrons produced in lead above and around the scintillators. The ZEPLIN-II veto system detected neutrons produced in lead below and around the liquid scintillator vessel. This and other differences in configuration coupled to possible inaccuracies in GEANT4 modelling of the angular distribution of neutron emission may explain the observed discrepancy between the two results.

We have also explored the evolution of the neutron production yield with successive versions of GEANT4. To do this further simulations of a mono-energetic $\mu^{-}$-beam focused on a lead block have been performed. Table~\ref{neutron_yield_table} summarises the results, including the yield obtained with version 8.2 from Ref.~\cite{Lindote}. In addition, combination of different physics lists and GEANT4 versions are listed, also linking the custom list used in~\cite{Lindote} to the current high energy reference lists. The bespoke physics list is very similar to {\tt QGSP$\_$BIC$\_$HP}, featuring the Quark-Gluon String (QGS) theoretical model at high energies coupled to nuclear de-excitation with a pre-compound model, the intra-nuclear Binary Cascade (BIC) model below 6~GeV and the data driven high precision neutron package ({\tt NeutronHP}) to transport neutrons below 20~MeV down to thermal energies. Reasonable variation of change-over energies between the BIC and QGS models in the custom physics list in comparison to the reference one has little impact ($<$3$\%$) on the overall neutron yield. A steady increase with every new version of GEANT4 is demonstrated.

\begin{table}
  \centering 
  \caption{Muon-induced production yields for neutrons for different versions of GEANT4 and physics lists (for 260~GeV muons). The neutron yield from version 8.2 is based on the value reported in Ref.~\cite{Lindote}. A small modification has been applied to correct for a previously unaccounted error in the rejection of neutrons produced in neutron inelastic processes to avoid double counting (referred to as `stars' in that work).}
      \vspace{2mm}
  \label{neutron_yield_table}
  \begin{tabular}{ p{1.4cm}p{2cm}p{4cm}}
  \hline
 GEANT4 version & physics list & muon-induced~neutron yield [neutrons/muon/(g/cm$^{2}$)]\\
 \hline
 \hline
 8.2 & custom list & (2.846$\pm$0.006) $\times$~10$^{-3}$ \\
 9.4 & custom list & (3.304$\pm$0.003)~$\times$~10$^{-3}$ \\
 9.4 & {\tt QGSP$\_$BIC$\_$HP} & (3.376$\pm$0.003)~$\times$~10$^{-3}$ \\
 9.4 & {\tt Shielding} & (3.682$\pm$0.003)~$\times$~10$^{-3}$ \\
 9.5 & {\tt QGSP$\_$BIC$\_$HP} & (3.993$\pm$0.004)~$\times$~10$^{-3}$ \\
 9.5 & {\tt QGSP$\_$BERT$\_$HP} & (4.369$\pm$0.004)~$\times$~10$^{-3}$\\
 9.5 & {\tt FTFP$\_$BERT} & (4.467$\pm$0.004)~$\times$~10$^{-3}$\\
 9.5 & {\tt Shielding} & (4.594$\pm$0.004)~$\times$~10$^{-3}$ \\
  \hline
\end{tabular}
\end{table}

\begin{figure}
\begin{center}
\includegraphics[width=0.36\textwidth]{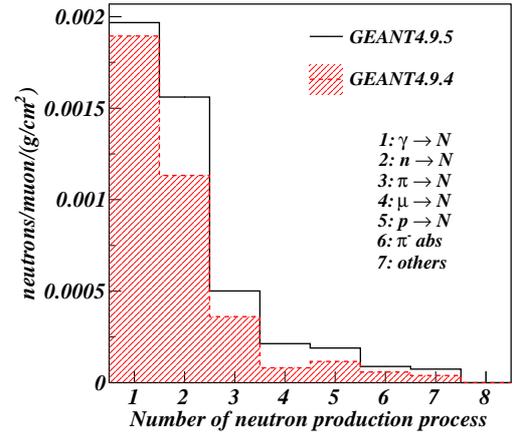}
\caption{(Colour online) Absolute neutron yields of the most important production processes for muon-induced neutrons generated from firing 260~GeV $\mu^{-}$ on lead using the {\tt Shielding} physics list and GEANT4 version 9.5 (black histogram) and version 9.4 (red dashed histogram). The neutron creation processes are: 1: photo-nuclear interaction of $\gamma$-rays ($\gamma$ $\rightarrow$ N), 2: neutron inelastic scattering (n $\rightarrow$ N), 3: pion spallation ($\pi$ $\rightarrow$ N), 4: muon spallation ($\mu$ $\rightarrow$N), 5: proton spallation (p $\rightarrow$ N), 6: pion absorption ($\pi^{-}$ abs) and 7: all other neutron production processes.}
\label{neutron_yield_in_lead_process}
\end{center}
\end{figure}

The {\tt Shielding} physics list shows not only the largest muon-induced neutron production yield in comparison to other reference lists, but is also subject to the highest increase in going from version 9.4 to 9.5. This is explored in detail in Fig.~\ref{neutron_yield_in_lead_process}, showing the individual contributions from the most important neutron creation processes for muons in lead. The main difference lies in the increased neutron production in inelastic scattering of hadrons and in particular neutrons. A $\sim$38$\%$ higher production yield for this process is observed.

Part of the increase observed between versions 9.4 and 9.5 of the toolkit (applicable to all standard lists used in this study) can be attributed to the muon-nucleus interaction model ({\tt G4VDMuonNuclearModel}); as in previous versions, this still relies on the Kokoulin mu-nuclear cross-sections~\cite{muoninteraction}, but the final state of the hadronic vertex is now replaced by a $\pi^{0}$ interacting further through the Bertini intra-nuclear cascade. The previous model ({\tt G4MuNuclearInteraction}) replaced the virtual photon with $\pi^{+/-}$ instead, which would then interact through the low/high energy parameterised models (LEP/HEP) --- these are known to yield fewer neutrons. There has also been increased neutron production in the FTF model, which may account for some of the enhanced yields in the {\tt Shielding} and {\tt FTFP$\_$BERT} lists; The addition of the Reggeon cascade~\cite{fritiof}, which can cause more nucleon secondaries, is a possible explanation, but further study is required~\cite{wright}.

\section{Conclusion}

\noindent For the development of future rare-event searches, especially in the context of direct dark matter experiments, accurate data on muon-induced neutron yields in several materials is of great importance, as is the ability to simulate these processes using modern Monte Carlo toolkits. Complex models inform the design of large and expensive shielding and veto systems around these experiments, as well as the interpretation of their data (background expectations). There exists significant uncertainty in the simulated muon-induced neutron rate, as evidenced by the steady variation in the total neutron yield with every new version of GEANT4 and physics list; experimental measurements have been likewise uncertain.

In this study, a dataset from 319 days of operation of the \mbox{ZEPLIN--III} anti-coincidence detector has been analysed for high energy cosmic-ray muons. The number of muon-induced neutrons has been evaluated by detecting delayed \mbox{$\gamma$-ray} signals following radiative captures. A muon flux in the Boulby Underground Laboratory of (3.75$\pm$0.09)~$\times$~10$^{-8}$~muons/s/cm$^{2}$  has been determined, consistent with and improving upon previous measurements. The muon-induced neutron detection rate was measured to be 0.346$^{+0.007}_{-0.016}$~neutrons/muon (quadratically combined statistical and systematic errors) traversing the ZEPLIN--III scintillator veto. Monte Carlo simulations, using GEANT4 (version 9.5) and the {\tt Shielding} physics list with the same cuts and thresholds applied as used for the analysis of the data, resulted in a neutron capture rate of 0.275$^{+0.005}_{-0.008}$~neutrons/muon, which is $\sim$20$\%$ lower than the experimentally measured value. However, absolute rates aside, the simulation reproduced very well all tested parameters, strengthening confidence in the results. The ratio of neutron rates between data and simulation have been used to evaluate a muon-induced neutron yield in pure lead of (5.78$^{+0.21}_{-0.28}$)~$\times$~10$^{-3}$~neutrons/muon/(g/cm$^{2}$) for a mean muon energy of 260~GeV. Additional simulations exploring previous versions of the GEANT4 simulation package confirm the trend of an increasing neutron production rate in lead with every successive distribution of GEANT4 (also shown in other simulation studies~\cite{muonsim,vito}).

Finally, our results confirm the very significant contribution of lead to the production of muon-induced neutrons. As such, the use of lead-based shielding to prevent $\gamma$-rays from the environment to propagate into the sensitive volume of the detector should be carefully assessed for any future rare event search. Alternative shielding compositions, such as large water tanks surrounding the detectors, are already being used in some current dark matter searches~\cite{lux,DEAP,XMASS}, as well as discussed for near future next generation experiments~\cite{LZ,eureca}.

\vspace{10mm}

\newproof{ack}{Acknowledgement}
\begin{ack}

The UK groups acknowledge the support of the Science \& Technology Facilities Council (STFC) for the ZEPLIN--III project (in particular, the analysis work presented here was supported by STFC grants ST/K006436/1,~ST/K003178/1,~ST/K003208/1,~ST/K006428/1, ST/K003186/1, ST/K006444/1 and ST/K006770/1) and for maintenance and operation of the underground Palmer Laboratory which is hosted by Cleveland Potash Ltd (CPL) at Boulby Mine, UK.  The project would not have been possible without the co-operation of the management and staff of CPL. Additionally, we want to thank the Boulby science facility team for their support during underground aspects of this work. We also acknowledge support from a Joint International Project award, held at ITEP and Imperial College, from the Russian Foundation of Basic Research (08-02-91851 KO\_a) and the Royal Society. LIP--Coimbra acknowledges financial support from Funda\c c\~ao para a Ci\^encia e Tecnologia (FCT) through the project-grants CERN/FP/109320/2009 and /116374/2010, and postdoctoral grants SFRH/BPD/27054/2006, /47320/2008, /63096/2009 and /73676/2010. Furthermore, we acknowledge the Edinburgh Compute and Data Facilities for accommodating the heavy usage of the Edinburgh computer cluster `Eddie' for this analysis.
This work was supported in part by SC Rosatom, contract contract $\#$H.4e.45.90.11.1059 from 10.03.2011. The University of Edinburgh is a charitable body, registered in Scotland, with registration number SC005336.
\newline

\end{ack}


\bibliographystyle{elsarticle-num}
\biboptions{sort&compress}
\bibliography{references_muon_induced_neutrons_Lea_Reichhart}


\end{document}